\documentclass[manuscript]{aastex63}

\usepackage{color}

\newcommand{\aff}{\langle |f_{i}| \rangle}
\newcommand{\maff}{\langle |f_{d}| \rangle}
\newcommand{\emix}{|\tilde {E}_{\rm mix}|}
\newcommand{\Lq}{L_{Q}}

\shorttitle{On measuring divergence}
\shortauthors{Gilchrist et al.}

\begin{document}

\title{On measuring divergence for magnetic field modeling}

\correspondingauthor{S.A. Gilchrist}
\email{sgilchrist@nwra.com}

\author{Gilchrist S.A.}
\affiliation{NorthWest Research Associates, \\
3380 Mitchell Lane, Boulder, \\
CO, 80301, USA}

\author{Leka K.D.}
\affiliation{NorthWest Research Associates, \\
3380 Mitchell Lane, Boulder, \\
CO, 80301, USA}

\author{Barnes G.}
\affiliation{NorthWest Research Associates, \\
3380 Mitchell Lane, Boulder, \\
CO, 80301, USA}

\author{Wheatland M.S.}
\affiliation{Sydney Institute for Astronomy, School of Physics, \\
The University of Sydney,\\
NSW 2006, Australia}

\author{DeRosa M.L.}
\affiliation{Lockheed Martin Solar and Astrophysics Laboratory\\
3251 Hanover St. B/252, Palo Alto,\\
CA, 94304, USA} 

\begin{abstract}
A physical magnetic field has a divergence of zero. Numerical error
in constructing a model field and computing the divergence,
however, introduces a finite divergence into these calculations.
A popular metric for measuring divergence is the 
average fractional flux $\aff$. We show that $\aff$ 
scales with the size of the computational mesh,
and may be a poor measure of divergence because it becomes
arbitrarily small for increasing mesh resolution, without the divergence actually decreasing. 
We define a modified version of this metric that
does not scale with mesh size. We apply the new metric
to the results of DeRosa et al. (2015), who measured $\aff$ for 
a series of Nonlinear Force-Free Field (NLFFF) models of the 
coronal magnetic field based on solar boundary data binned 
at different spatial resolutions. 
We compute a number of divergence metrics for the DeRosa et al. (2015) data and analyze the 
effect of spatial resolution on these metrics using a non-parametric
method. We find that some of the trends reported by DeRosa et al.
(2015) are due to the intrinsic scaling of $\aff$. We also find
that different metrics give different results for the same data set 
and therefore there is value in measuring divergence via several metrics. 
\end{abstract}

\section{Introduction} 
\label{sect_intro}

The solar coronal magnetic field is difficult to directly infer, and so it has 
become common to rely on Nonlinear Force-Free magnetic Field (NLFFF)
``extrapolations'' to study it. An NLFFF extrapolation uses
observations of the vector magnetic field at the photosphere
to construct a three-dimensional model of the coronal magnetic 
field. The coronal extrapolation problem has 
a long history and is the subject of several reviews 
\citep{1989SoPh..120...19A,2012LRSP....9....5W,2013SoPh..288..481R}.

A magnetic field is force-free if it satisfies the nonlinear force-free equations \citep{sturrock1994plasma}:
\begin{equation} 
  (\nabla\times\mathbf B)\times\mathbf B = 0,
  \label{equ_ff}
\end{equation}
and
\begin{equation}
  \nabla\cdot\mathbf B = 0.
  \label{equ_solen}
\end{equation}
A force-free magnetic field is a natural 
equilibrium state for a magnetized plasma where gas pressure and other forces are 
negligible --- the equilibrium being one where the magnetic
Lorentz force is self balancing. It is also the minimum energy state
for a specified connectivity of field lines \citep{1989SSRv...51...11S}.
An extrapolation involves solving
Equations (\ref{equ_ff})-(\ref{equ_solen}) in a three-dimensional volume subject to
boundary conditions on the bottom boundary derived from spectro-polarimetric observations 
of the photospheric magnetic field. Such boundary data are generally
noisy and are additionally inconsistent with the force-free model, because there
are significant gas pressure and gravity forces at the photosphere
\citep{1995ApJ...439..474M,2001SoPh..203...71G}. This can cause
problems for the modeling. The extrapolated magnetic field model
may have residual forces and a finite divergence \citep{2009ApJ...696.1780D,2015ApJ...811..107D}.\\

Violations of the solenoidal condition are a particular problem for NLFFF modeling
because they lead to nonphysical magnetic fields. They may also lead to spurious
estimates for the magnetic energy \citep{2013A&A...553A..38V,2015ApJ...811..107D}, 
and accurate estimates of energy are often a goal of NLFFF 
modeling (e.g. \citealp{2008A&A...484..495T}).

Consequently, it is important to measure $\nabla\cdot\mathbf B$ for NLFFF models
to properly interpret the results. There are many ways of doing this. 
The volume integral of either $|\nabla\cdot\mathbf B|$ or 
$|\nabla\cdot\mathbf B|^2$ is a common measure
(e.g. \citealp{2012AJ....144...33T,2006SoPh..235..161S}). 
The non-solenoidal contribution to the magnetic energy is another measure 
\citep{2013A&A...553A..38V,2014SoPh..289.4453M,2014ApJ...788..150S,2018SoPh..293..130M}. 
\citet{2018SoPh..293..130M} consider the total signed magnetic flux over the boundary
in addition to the non-solenoidal component of the energy.
This list of metrics is not exhaustive, but demonstrates that there are a variety of ways
of measuring $\nabla\cdot\mathbf B$ that are in use; each has different strengths 
and weaknesses. The total signed magnetic flux is only sensitive to 
the volume integral of $\nabla\cdot\mathbf B$, which may vanish despite
local non-zero values of $\nabla\cdot\mathbf B$ that cancel in the integral
due to contributions within the volume with different signs.
The nonsolenoidal magnetic energy is strictly only uniquely defined
if the net $\nabla\cdot\mathbf B$ is zero. However,  \citet{2013A&A...553A..38V}
show that this may not be a serious problem in practice.\\ 

The average fractional flux, $\aff$, is a commonly-used measure of the divergence of a 
vector field. It was first defined by  \citet{2000ApJ...540.1150W}
and is zero for a perfectly solenoidal vector field. In principle,
it can measure the divergence of any vector field, however it is primarily 
used to measure $\nabla\cdot\mathbf B$ in the context of modeling solar 
magnetic fields. We argue that $\aff$ is generally unsuitable as a metric for measuring
$\nabla\cdot\mathbf B$ because it scales with mesh  resolution 
independently of $\nabla\cdot\mathbf B$. We show in Section \ref{sec_metric}
that 
\begin{equation}
  \aff \sim (\Delta V)^{\frac{1}{3}},
\end{equation}
for a magnetic field defined on a mesh where each cell has 
uniform volume $\Delta V$. The symbol $\sim$, in this context 
indicates an asymptotic, scaling relationship. The tendency
for $\aff$ to scale with mesh resolution is mentioned by 
\citet{2013A&A...553A..38V}, who notes that values of $\aff$
may only be strictly compared between meshes with the same cell volume.
This makes it difficult to use $\aff$ for comparisons between
different studies, as differences in $\aff$ may only reflect
differences in the mesh spacing.\\

Why does it matter if the particular metric, $\aff$, is potentially
a poor measure of $\nabla\cdot\mathbf B$?
There are two reasons. The first reason is that $\aff$ is popular. Indeed,
a cursory survey of the literature indicates that at least 
50 papers published over the last 20 years have used $\aff$.  On average,
this is about 2-3 papers per year. Our survey considered only papers that 
directly cited \citet{2000ApJ...540.1150W}, so the actual number is 
likely higher. 

The second reason for considering $\aff$ is that it was recommended
by a well-cited NLFFF workshop paper \citep{2015ApJ...811..107D}.
Based on International Space Science Institute (ISSI)
workshops held in 2013 and 2014, \citet{2015ApJ...811..107D} 
considered the effect of spatial resolution on $\nabla\cdot\mathbf B$ in NLFFF modeling.
They constructed NLFFF models for NOAA active region AR 10978 using
different numerical NLFFF methods and spatial resolutions.
For the models considered, \citet{2015ApJ...811..107D}  found
that $\aff$ tends to decrease as spatial resolution is increased 
($\Delta V$ becomes smaller). The decrease in $\aff$ was interpreted as a true 
decrease in $\nabla\cdot\mathbf B$, however, given that $\aff$ scales with 
$(\Delta V)^\frac{1}{3}$, this interpretation may be called into question.

In this paper we have two aims. Firstly, we aim to 
give a formal description of the scaling problem for 
$\aff$ and to propose a new metric  --- which we call the modified
fractional flux $\maff$ --- that is based on $\aff$, but is free from 
the scaling problem. The scaling problem and the new metric are introduced in 
Section \ref{sec_metric}. Secondly, we revisit the question of \citet{2015ApJ...811..107D} using
additional metrics not considered by \citet{2015ApJ...811..107D}, including
$\maff$. In Section \ref{sec_derosa} we summarize the study
of \citet{2015ApJ...811..107D}. In Section  \ref{sec_issi} we present the results
of the metrics applied to the \citet{2015ApJ...811..107D} data, and
in Section \ref{sec_calc_metrics} we perform a non-parametric trend analysis
of these metric data to ascertain the effect of spatial resolution.
We also examine the concordance between different metrics, 
i.e. if we rank solutions using different metrics, to what extent
do these rankings agree/differ. We again address this problem statistically. 
In Section \ref{sec_con} we discuss the results and present the conclusions.


\section{The average fractional flux and the scaling problem}
\label{sec_metric}

The average fractional flux is defined by \citet{2000ApJ...540.1150W}
as 
\begin{equation}
  \langle |f_i| \rangle = \left \langle                      
                      \frac{|\int_{\partial S_i} \mathbf B\cdot d \mathbf S|}
                      {\int_{\partial S_i} |\mathbf B|dS }
                      \right \rangle,
  \label{equ_f1}                      
\end{equation}
where $\partial S_i$ is the surface of a voxel whose
volume is $S_i$. The subscript $i$ is used to indicate that the 
computational mesh on which $\mathbf B$ is defined is comprised
of many such voxels. The operator $\langle \rangle$ is the arithmetic mean over 
every voxel of the mesh. The ratio is the total flux over 
each voxel normalized by the average of $|\mathbf B|$ over the surface bounding the voxel.
The surface integral in the numerator is related to
$\nabla\cdot\mathbf B$ by Gauss's law, i.e.
\begin{equation}
  \int_{\partial S_i} \mathbf B\cdot d \mathbf S = 
  \int_{S_i} (\nabla\cdot\mathbf B)_i dV.
\end{equation}
It follows that if $\nabla\cdot\mathbf B = 0$, then $\aff =0$ too.

The most common form of $\aff$ that appears in the literature is for
a uniform Cartesian mesh:
\begin{equation}
  \aff = \Delta x \left\langle
                      \frac{|\nabla\cdot\mathbf B|_i}{6 |\mathbf B|_i}
                      \right\rangle,
  \label{equ_f2}                      
\end{equation}
where $\Delta x$ is the spacing of the mesh. In deriving this
form, the integrals in Equation (\ref{equ_f1}) are approximated by
cell and face averages \citep{2000ApJ...540.1150W}, meaning 
Equation (\ref{equ_f2}) agrees with Equation (\ref{equ_f1}) only
to within some truncation error in $\Delta x$. However, nothing that we present
here depends critically on this approximation.

Equation (\ref{equ_f2}) is a product of $\Delta x$ and an average
term, which strongly suggests $\aff \sim \Delta x$. However, it is also important to realize that
the average term depends on $\Delta x$ too: at different resolutions, 
the average is performed over different samplings of 
$|\nabla\cdot\mathbf B|/|\mathbf B|$, so that
even if $|\nabla\cdot\mathbf B|/|\mathbf B|$ is independent of $\Delta x$,
the average will have some $\Delta x$ dependence. As a function
of resolution, the term $\langle |\nabla\cdot\mathbf B|/|\mathbf B| \rangle$
is a set of partial sums that 
converges to a limit at a rate that depends on $\Delta x$.
In particular, it can be shown that, generally
\begin{equation}
  \aff = \frac{\Delta x}{6V}\int_{S}\frac{|\nabla\cdot\mathbf B|}{|\mathbf B|} dV
  + \mathcal{O}(\Delta x^2),
  \label{equ_flim}
\end{equation}
where $S$ is the whole domain with volume $V$. The ``error'' term
$\mathcal{O}(\Delta x^2)$ reflects the fact that the average
is a Riemann sum that generally differs from the integral by a truncation error
of order $\mathcal{O}(\Delta x)$. The coefficient of the average 
is $\Delta x$, leading to the term $\mathcal{O}(\Delta x^2)$ in 
Equation (\ref{equ_flim}). Hence, we may conclude that,
generally, $\aff \sim \Delta x$ to first order in $\Delta x$.

The tendency of $\aff$ to decrease with resolution is not a special consequence of a uniform
mesh. Equation (\ref{equ_f1}) involves the ratio of a volume integral to a surface integral.
The numerator has scaling $\sim \Delta V_i$, where $\Delta V_i$ is the volume of the voxel $i$,
while the denominator has scaling $\sim \Delta V_i^{2/3}$. Hence 
the ratio has scaling $\sim \Delta V_i^{1/3}$, and therefore
$\aff \sim \langle \Delta V_i^{1/3} \rangle$.
The quantity $\Delta V_i^{1/3}$ has the form of an ``effective'' linear 
dimension of each voxel, i.e. it is the side length of a cube with
the same volume. 

To address this issue with $\aff$, we now introduce the modified fractional flux, defined as
\begin{equation}
  \maff = \left \langle
                      \frac{|\int_{\partial S_i} \mathbf B\cdot d \mathbf S|}
                      {\int_{ S_i} |\mathbf B|dV }
                      \right \rangle.
  \label{equ_maff}                      
\end{equation}
Equation (\ref{equ_maff}) differs from Equation (\ref{equ_f1})
in that the denominator is the integral over volume, i.e.
$\maff$ involves a different normalization of the net flux
at each voxel. On a uniform mesh, 
\begin{equation}
  \maff = \frac{6}{\Delta x} \aff.
  \label{equ_maff_u}
\end{equation}
Unlike $\aff$, which is non-dimensional, $\maff$ has units of inverse length.
As a result, values of $\aff$ and $\maff$ cannot be directly  
compared because they are in different units. In principle, one could
non-dimensionalize $\maff$ with some characteristic length scale, e.g. 
$V^{1/3}$. 

We expect
\begin{equation}
  \maff \sim c + \mathcal{O}(\Delta V^{1/3}),
\end{equation}
where $c$ depends on $\nabla\cdot\mathbf B$, but is 
independent of $\Delta V$. This is an improvement over $\aff$, because $\maff$ has a finite 
limit for small $\Delta V^{1/3}$ when $\nabla\cdot\mathbf B \ne 0$.

\subsection{Application of $\aff$ and $\maff$ to a test case}

In this section we apply $\aff$ and $\maff$ to a simple test
case that demonstrates the scaling problem for $\aff$, and we
show that $\maff$ is free from this problem. In this section
we use non-dimensional units. Magnetic fields, lengths, 
and differential operators are scaled by an unspecified 
characteristic magnetic field strength $B_{\rm c}$
and length scale $L_{\rm c}$. Non-dimensional quantities are 
indicated with bars, e.g. the non-dimensional mesh
spacing is $\overline{\Delta x} = \Delta x/L_{\rm c}$. The actual
dimensions are not important to the results.

Consider the magnetic field
\begin{equation}
  \overline{\mathbf B} = \overline{B_{\rm s}} \overline{z} \mathbf{ \hat z},
  \label{equ_example}
\end{equation}
where $\mathbf {\hat z}$ is a the Cartesian unit vector,
and $\overline{B_{\rm s}}$ is a constant that sets the magnitude of the field
in non-dimensional units.
The divergence is 
\begin{equation}
  \overline{\nabla}\cdot \overline{\mathbf B} = \overline{B_{\rm s}}. 
\end{equation}
In what follows, we take $\overline{B_{\rm s}} = 1/2$, 
and a uniform mesh with spacing $\overline{\Delta x}$ that spans a
three-dimensional box $(\overline{x},\overline{y},\overline{z})\in
[1,2]\times[1,2]\times[1,2]$. For these parameters, it can be shown that
\begin{equation}
  \aff = \frac{\overline{\Delta x}\log(2)}{6} + \mathcal{O}(\overline{\Delta x^2}),
\end{equation}
and
\begin{equation}
  \maff L_{\rm c}  = \log(2) + \mathcal{O}(\overline{\Delta x}).
\end{equation}
The results are independent of $B_{\rm s}$.
To first order $\aff \sim \overline{\Delta x}$ and $\maff \sim 1$.
The higher-order terms are the result of the convergence of
the average term.

\begin{figure}[!h]
  \plottwo{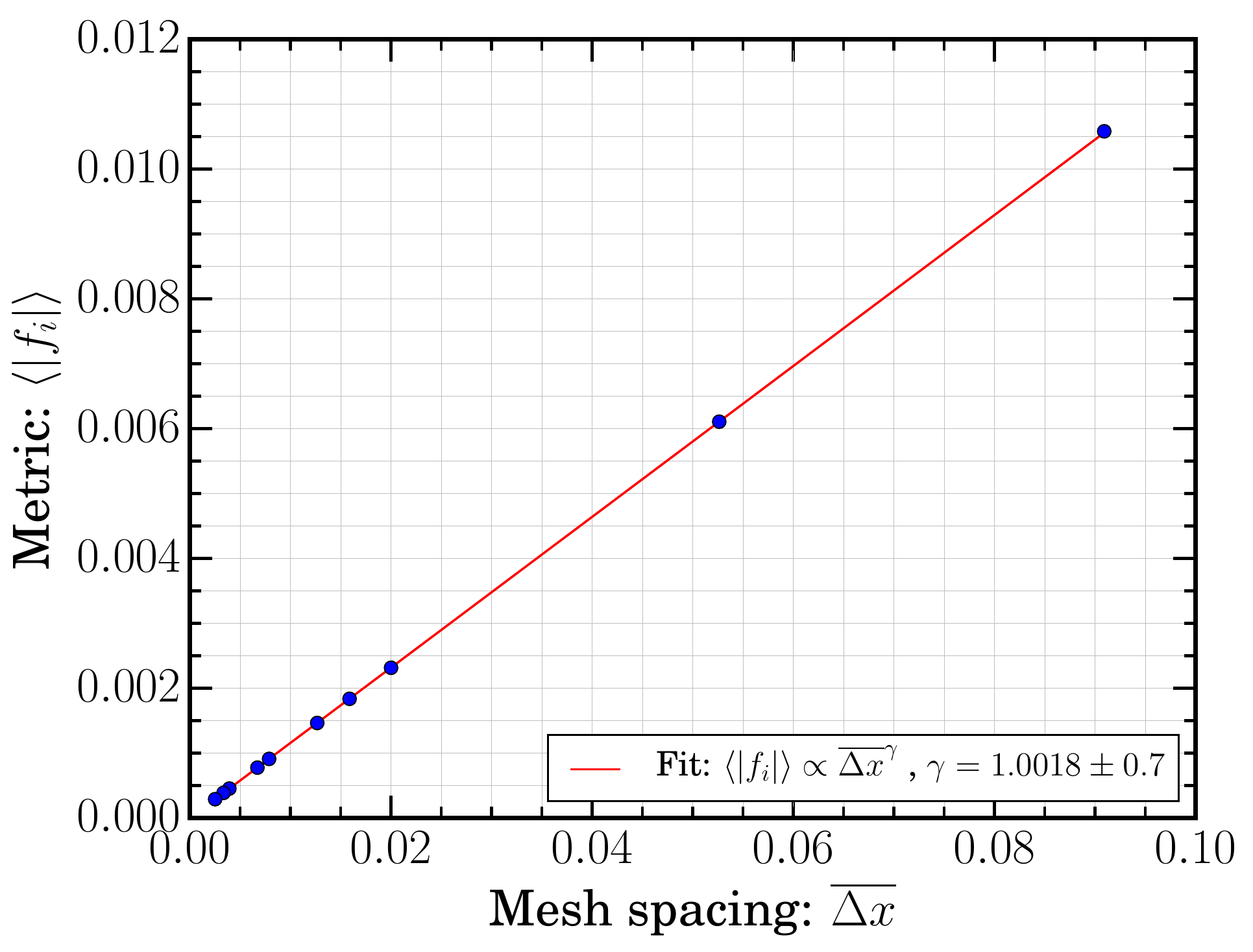}{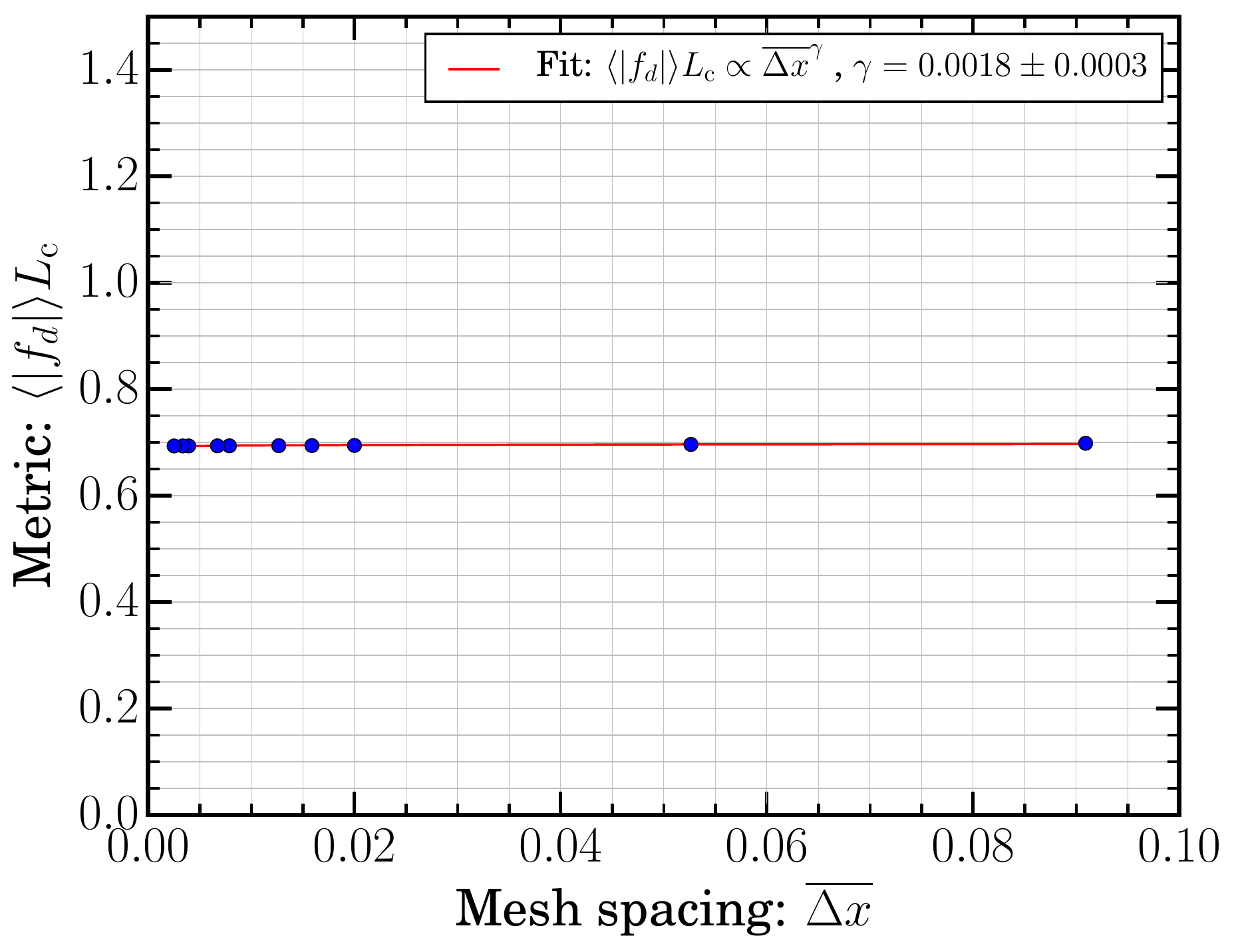}
  \caption{The left panel shows $\aff$ computed
  for the example field defined by Equation (\ref{equ_example})
  on a series of uniform meshes with grid spacing $\overline{\Delta x}$. The 
  solid line is a power-law fit to the data with index 
  $\gamma \approx 1$. The metric decreases as $\overline{\Delta x}$ decreases. 
  The right panel shows 
  $\maff$ computed for the same magnetic field
  and the same meshes. The solid line is a power-law fit 
  with index $\gamma \approx 0$. The metric $\maff$
  remains approximately constant as a function of resolution.
  In these examples, the length is measured in terms of an unspecified
  characteristic length $L_{\rm c}$, which is constant. 
  The magnitude of the two metrics differ, so a direct comparison is not meaningful
  because $\maff$ is a dimensional quantity, while $\aff$ is not. 
  In non-dimensional units $\maff$ depends on the arbitrary scaling $L_c$, 
  whereas $\aff$ does not.}  
  \label{fig_example}
\end{figure}

The left panel of Figure \ref{fig_example} shows $\aff$ for this field 
computed at different resolutions. The extent of the volume 
is unchanged with resolution. The metric, $\aff$, 
decreases systematically with increasing spatial resolution, 
however the underlying divergence is unchanged. The solid line is a 
power-law fit to the data with power-law index $\gamma = 1.0018 \pm 0.0004$, with the 
uncertainty derived from the covariance matrix of the fit ---
the power-law index, $\gamma$, is close to unity, but not exactly, 
because the scaling of $\aff$ departs from $\sim \overline{\Delta x}$ for
large $\overline{\Delta x}$.

The right panel shows $\maff$ computed under
the same conditions. The $\maff$ metric changes very 
little with resolution. A power law fit has index 
$\gamma =  0.0018\pm 0.0004$, with the 
uncertainty derived from the covariance matrix of the fit.
Again, $\gamma$ is close to the asymptotic value of zero, but departs
from constant scaling for large values of $\overline{\Delta x}$.


\section{Description of DeRosa et al. (2015) study}
\label{sec_derosa}

\citet{2015ApJ...811..107D} performed NLFFF extrapolations of NOAA active region
AR 10978 on 2007 December 13 using magnetic field boundary conditions 
derived from the {\it Hinode}/Solar Optical Telescope
Spectro-Polarimeter ({\it Hinode}/SOT) observations
\citep{2008SoPh..249..167T,2013SoPh..283..579L}. A set of
boundary conditions for the modeling was constructed 
at different spatial resolutions by binning the {\it Hinode}/SOT Stokes spectra by 
different integer factors. The binned spectra were then subject to
spectro-polarimetric inversion, ambiguity resolution, and 
remapping to a flat heliographic tangent plane suitable for
computing the NLFFF extrapolations in Cartesian coordinates.
The coordinate mesh on the tangent plane had uniform spacing.
A complete description of the data preparation is given 
in \citet{2015ApJ...811..107D}.

Table \ref{table_metrics} lists the models and shows the bin factors used in
the \citet{2015ApJ...811..107D}  study. A bin factor of unity
corresponds to no binning, but is not shown because \citet{2015ApJ...811..107D}
did not perform extrapolations at this resolution due 
to the high computational intensity of the calculations. For a bin factor of unity,
the mesh scale is $0.106\,{\rm Mm}$. The mesh spacing for the other bin factors is given by
$0.106\,{\rm Mm}$ multiplied by the relevant bin factor.

\citet{2015ApJ...811..107D} used five different NLFFF extrapolation
methods: the optimization method (OPTI) described in \citet{2010A&A...516A.107W}
and \citet{2012SoPh..281...37W}; the magnetofrictional method (MAGF) described in \citet{2007SoPh..245..263V,2010A&A...519A..44V}; and three codes based on 
different implementations of the Grad-Rubin method \citep{24756}, namely 
CFIT \citep{2007SoPh..245..251W}, XTRAPOL \citep {2006A&A...446..691A,2010A&A...522A..52A} and FEMQ 
\citep{2006A&A...446..691A}. 

Boundary conditions for each NLFFF method were derived from the binned
{\it Hinode} vector magnetogram data using an approach specific 
to each method. Different approaches
to smoothing, censoring, and pre-processing were applied to 
derive boundary conditions in each case. The detailed methods
are described in \citet{2015ApJ...811..107D}.

All the extrapolations were performed in Cartesian coordinates.
Some methods used a nonuniform mesh for the calculation,
but for the analysis, all data were mapped to a uniform Cartesian
mesh: the spacing and extent of the mesh were consistent across methods.
The spacing differed between bin factors but the extent of the domain
was the same in each case.\\

For methods based on the Grad-Rubin iteration, there are two 
solutions for each bin factor, labeled $P$ and 
$N$. For the Grad-Rubin method, boundary conditions on the electric
current density are only prescribed on one polarity of the normal
component of the magnetic field $B_n$ \citep{24756}. Therefore, two solutions
are possible given one set of boundary data. For the $P$ solution, 
electric current is prescribed at points where $B_n > 0$, and
for the $N$ solution, electric current is prescribed at points where $B_n < 0$. 

We note that all of the authors of the current work were
involved in the \citet{2015ApJ...811..107D} study. S.A. Gilchrist
and M.S. Wheatland computed the CFIT solutions.


\section{Analysis of DeRosa et al. (2015) results}
\label{sec_issi}

In this section we analyze the magnetic field data of 
\citet{2015ApJ...811..107D} using four metrics: 
the average fractional flux $\aff$, the modified fractional flux $\maff$,
the total unsigned divergence $L_Q$, and the mixed component of
the non-solenoidal energy $\emix$. Two of these metrics have already been discussed 
in Section \ref{sec_metric}, and we describe the other two 
in Section \ref{sec_other_metrics}. In Section \ref{sec_calc_metrics}
we apply the four metrics to the data of \citet{2015ApJ...811..107D},
and we perform a non-parametric analysis of the metric data.


\subsection{Other measures of $\nabla\cdot\mathbf B$}
\label{sec_other_metrics}

One popular class of metric measures the 
total/average unsigned $\nabla\cdot\mathbf B$ in the volume.
One example from this class is the volume-averaged absolute divergence, given
by
\begin{equation}
  L_{Q} = \frac{1}{V}\int |\nabla\cdot\mathbf B| dV,
  \label{equ_Lq}
\end{equation}
which has units of flux per unit volume ($\rm Mx/cm^{3}$ in cgs).
Minor variations of this metric exist. For example, 
$|\nabla\cdot\mathbf B|^2$ may be used in place of 
$|\nabla\cdot\mathbf B|$ \citep{2006SoPh..235..161S,2012AJ....144...33T}. 
The integral may also be replaced by a discrete average \citep{2012RAA....12..563F}.
It is important to note that $L_{Q}$ depends on the magnitude of  
$\mathbf B$, which complicates comparisons between cases where
the characteristic magnetic field strength differs. For example, comparisons between
extrapolations of different active regions whose magnetic field strengths differ
significantly cannot be strictly compared without correcting for this difference in some way. 
We consider this type of metric because of its prevalence in the literature.\\

As another metric, we consider the non-solenoidal energy metric of \citet{2013A&A...553A..38V}.
A magnetic field can be decomposed into solenoidal and non-solenoidal components:
\begin{equation}
  \mathbf B = \mathbf B_{\rm p,s} + \mathbf B_{J,s} + \nabla\zeta + \nabla\psi,
\end{equation}
where $\mathbf B_{\rm p,s}$ and $ \mathbf B_{J,s} $ are the non-solenoidal components,
and $\nabla\zeta$ and $\nabla\psi$ are the solenoidal components. We define 
these in more details below, here it suffices to say that both 
$\nabla\zeta$ and $\nabla\psi$ are zero when $\nabla\cdot\mathbf B = 0$.
The total magnetic energy of the field is then
\begin{equation}
  E = E_{\rm p,s} + E_{J, \rm s} + E_{\rm p,ns} +  E_{J, \rm ns} + E_{\rm mix},
  \label{equ_E_decomp23}
\end{equation}
where
\begin{equation}
  E_{\rm p,s} = \frac{1}{8\pi}\int_V B^2_{p,s} dV,
\end{equation}
\begin{equation}
  E_{J, \rm s} = \frac{1}{8\pi}\int_V B^2_{J, \rm s} dV
\end{equation}
\begin{equation}
  E_{\rm p,ns} = \frac{1}{8\pi}\int_V |\nabla\zeta|^2 dV,
\end{equation}
\begin{equation}
  E_{J, \rm ns} = \frac{1}{8\pi}\int_V |\nabla\psi|^2 dV,
\end{equation}
and
\begin{eqnarray}
  E_{\rm mix} & = & \frac{1}{4\pi} \left ( \int_V \mathbf B_{\rm p,s} \cdot \nabla \zeta  dV  + \int_V \mathbf B_{J,\rm s} \cdot \nabla \psi  dV \right. \\
        & + & \int_V \mathbf B_{\rm p,s} \cdot \nabla \psi   dV + \int_V \mathbf B_{J,\rm s} \cdot \nabla \zeta dV \\
        & + & \left. \int_V \nabla \zeta \cdot \nabla \psi          dV + \int_V \mathbf B_{\rm p,s} \cdot  \mathbf B_{J,\rm s} dV \right ).
\end{eqnarray}
The vector fields used in the decomposition are constructed by first
splitting $\mathbf B$ into the sum
\begin{equation}
  \mathbf B = \mathbf B_{\rm p} + \mathbf B_{J},
  \label{equ_decomp1}
\end{equation}
where $\mathbf B_{\rm p}$ is the potential field that matches the normal component of 
$\mathbf B$ on the boundary, and $\mathbf B_{J}$  is
defined by Equation (\ref{equ_decomp1}), i.e.  $\mathbf B_{J} = \mathbf B - \mathbf B_{\rm p}$.
The field $\mathbf B_J$ is sometimes called the current-carrying component (e.g. \citealp{2015ApJ...811..107D}),
however this is misleading. It is more accurate to say that $\mathbf B_J$ is the field whose 
curl matches $\mathbf B$. It is important to note that even when the electric current density is 
zero everywhere, $\mathbf B_J$ will generally have a finite value, unless $\nabla\cdot\mathbf B = 0$.
The potential field $\mathbf B_{\rm p}$ is further decomposed into the sum 
\begin{equation}
  \mathbf B_{\rm p} = \mathbf B_{\rm p,s} + \nabla\zeta
  \quad \mbox{where}\quad
  \left \{
  \begin{array}{l}
    \nabla^2\zeta = \nabla\cdot\mathbf B_{\rm p} \\
    \left. \partial_n \zeta \right |_{\partial V} = 0,
  \end{array}
  \right .
\end{equation}
and where $\mathbf B_{\rm p,s}$ is the solenoidal component of $\mathbf B_{\rm p}$,
$\nabla\zeta$ is the non-solenoidal component of  $\mathbf B_{\rm p}$, $\partial_n$
is the normal derivative at the boundary, and $\nabla^2$ is the Laplace operator.
The $\mathbf B_J$ component is also decomposed into a sum of solenoidal
and non-solenoidal components:
\begin{equation}
  \mathbf B_{J} = \mathbf B_{ J,\rm s} + \nabla\psi
  \quad \mbox{where}\quad
  \left \{
  \begin{array}{l}
    \nabla^2\psi = \nabla\cdot\mathbf B_{ J} \\
    \left. \partial_n \psi \right |_{\partial V} = 0.
  \end{array}
  \right .
\end{equation}
\citet{2015ApJ...811..107D} found that $|E_{\rm mix}|$ was the largest
magnitude term in Equation (\ref{equ_E_decomp23}), for the
NLFFF models considered. This mixed term is a coupling energy between the solenoidal and 
non-solenoidal components of the magnetic field. As a metric for $\nabla\cdot\mathbf B$, we consider 
the non-dimensional form of $|E_{\rm mix}|$ defined by \citet{2013A&A...553A..38V} as
\begin{equation}
  \emix = \frac{| E_{\rm mix}|}{E},
\end{equation}
where $E$ is the total energy in the magnetic field $\mathbf B$. When
$\nabla\cdot\mathbf B = 0$ it follows that $\emix = 0$.


\subsection{Application of metrics to DeRosa et al. (2015) data}
\label{sec_calc_metrics}

We compute $\aff$, $\maff$, and $L_Q$ from the \citet{2015ApJ...811..107D}
data cubes\footnote{The  \citet{2015ApJ...811..107D} data
are available online at \url{https://doi.org/10.7910/DVN/7ZGD9P}.}.
We compute $\nabla\cdot\mathbf B$ using a centered-difference approximation
to the derivative \citep{Press:2007:NRE:1403886}.
This is consistent with the method used to compute $\nabla\cdot\mathbf B$
for analysis in \citet{2015ApJ...811..107D}, but is not necessarily
consistent with the numerical schemes used internally by the various NLFFF
methods/codes.  We do not recompute $\emix$. Instead, we simply rely on the value
from Table 4 of \citet{2015ApJ...811..107D}. 

\begin{figure}[!h]
  \plottwo{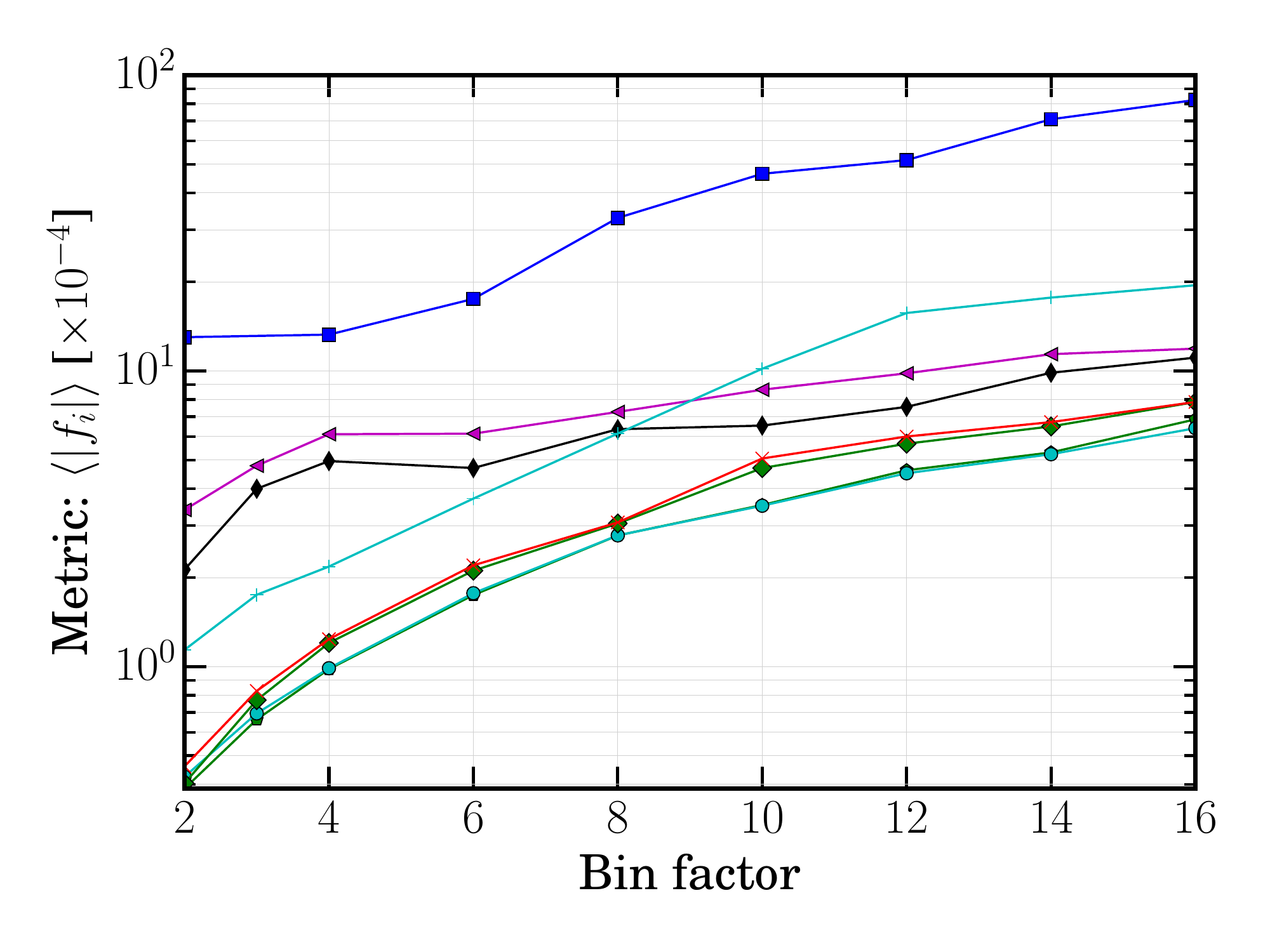}{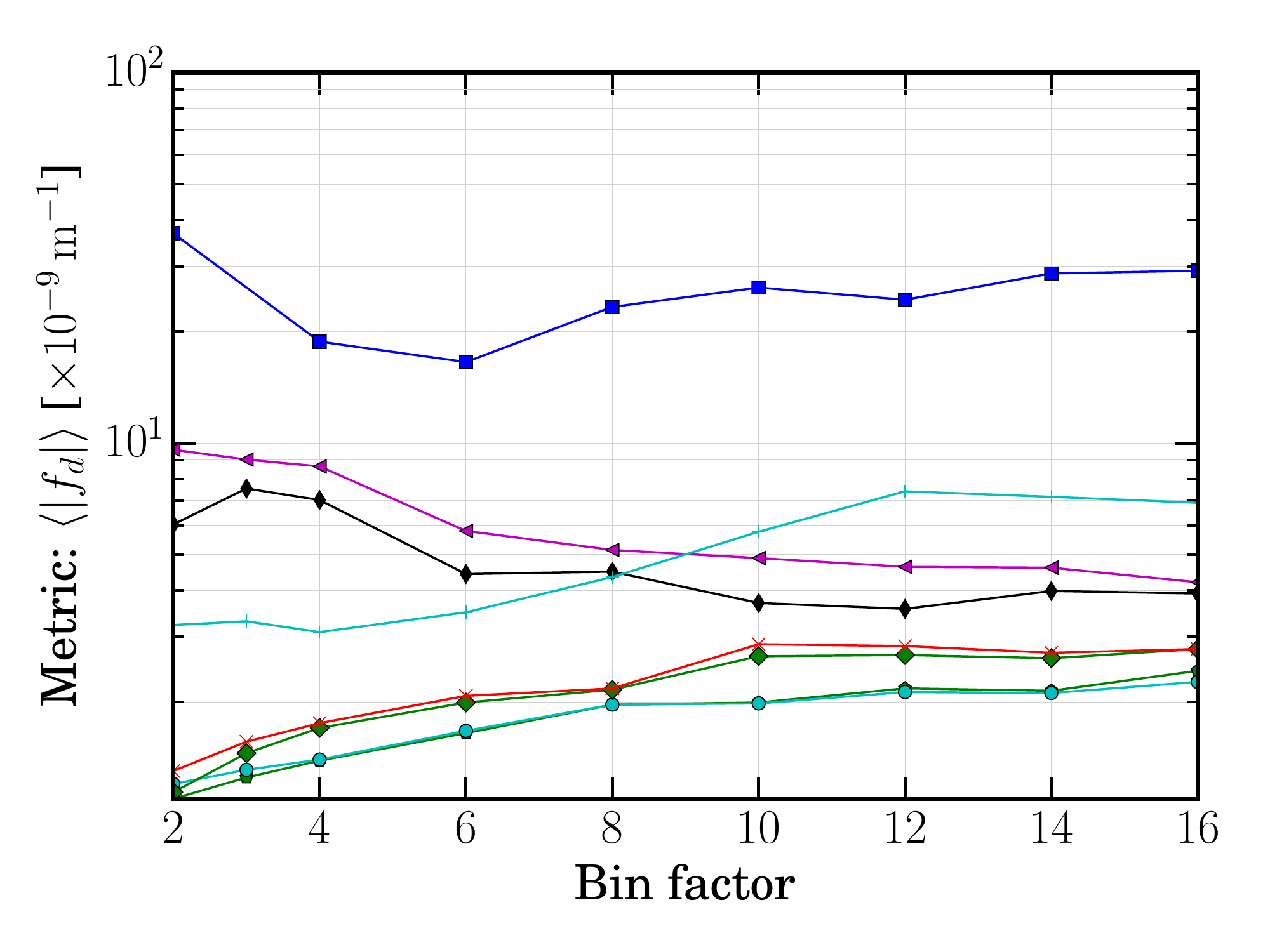}
  \plottwo{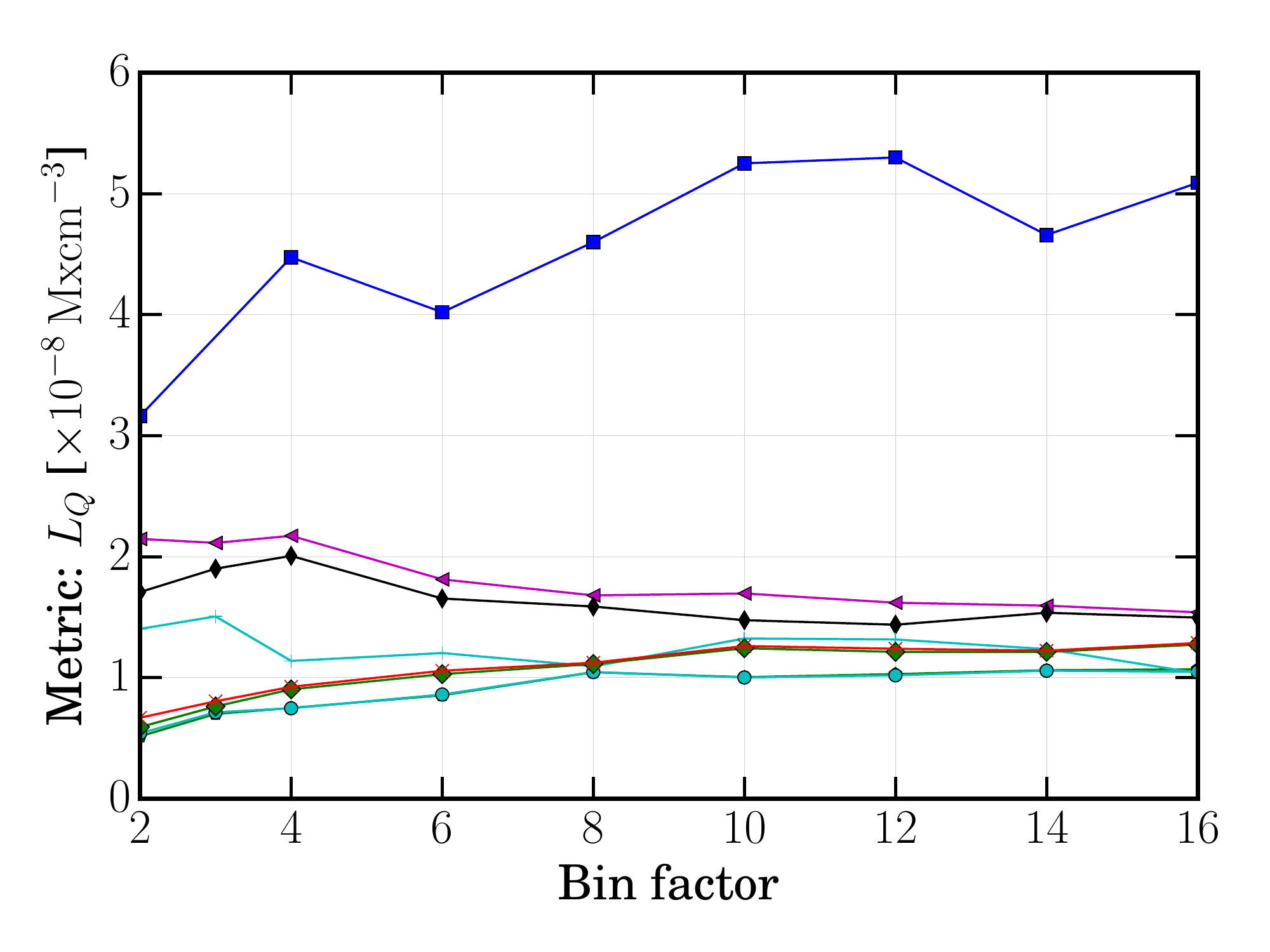}{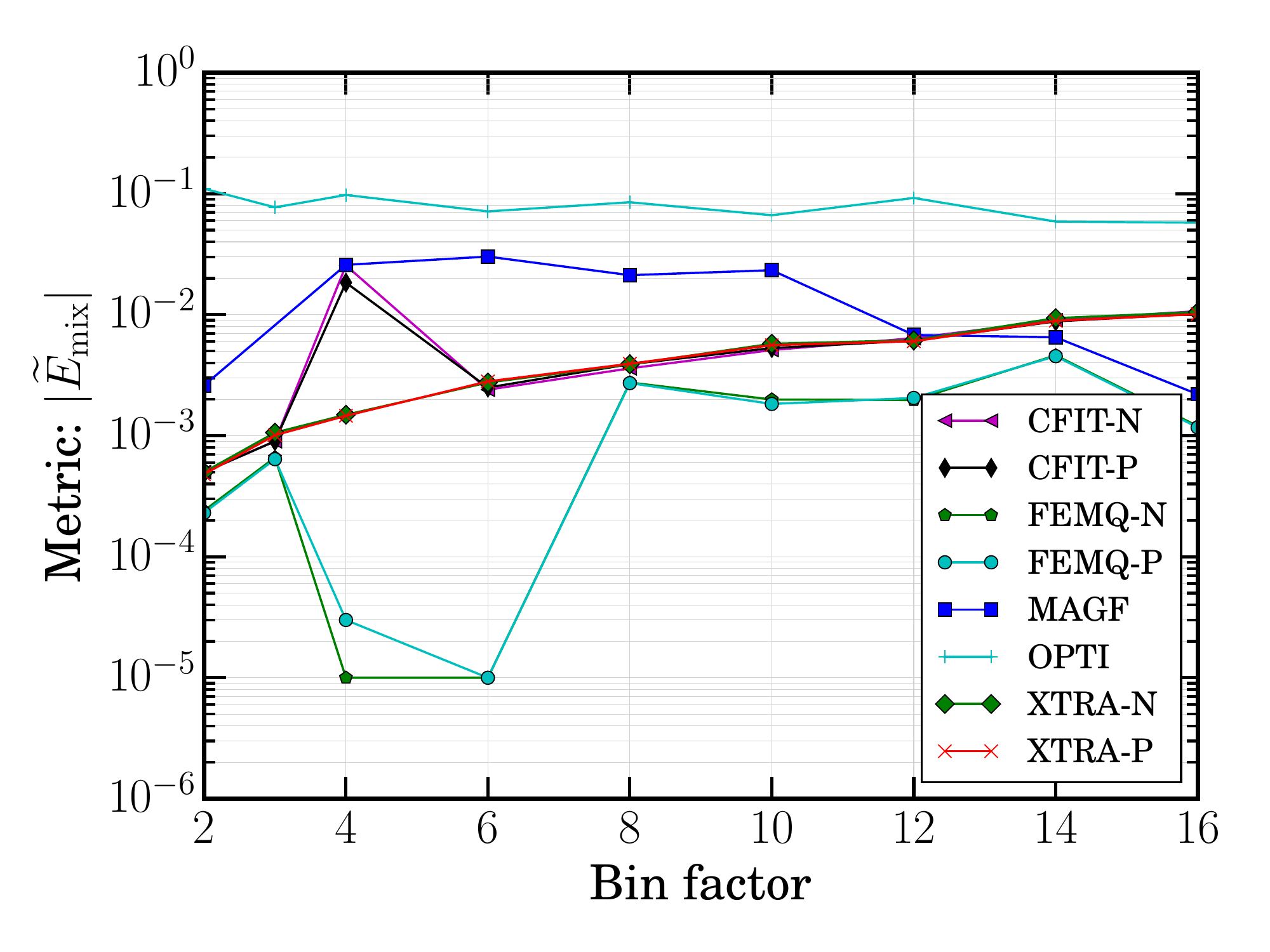}
  \caption{Metrics for evaluating $\nabla\cdot\mathbf B$ versus bin factor
  for the \citet{2015ApJ...811..107D} data. Lower bin factors correspond 
  to higher resolution. The numerical values are shown in Table \ref{table_metrics}. 
  The trend clearly visible in $\aff$ is not consistently evident. There are similar
  trends for some methods for some metrics.}
  \label{fig_metrics}
\end{figure}

Table \ref{table_metrics} shows $\aff$, $\maff$, $L_Q$, and 
$\emix$ for each method and bin factor. These results are 
shown also in Figure \ref{fig_metrics}. A visual inspection
of the data does not reveal any clear trend that is common 
to all methods.\\

\subsection{Rank correlation trends for metrics}
\label{sec_rank}

For each of the methods, we compute the Kendall $\tau$
rank-correlation coefficient \citep{kendall1962rank,daniel1978applied,
Press:2007:NRE:1403886} between the bin factor and each metric. Kendall's $\tau$ measures
the agreement (concordance) between two methods of ranking data. 
For two sets of data $x_i$ and $y_j$, Kendall's $\tau$ is defined as
\citep{kendall1962rank}
\begin{equation}
  \tau = \frac{2}{n(n-1)} \sum_{i=1}^n \sum_{j>i}^n \mbox{sgn}(x_i - x_j)\mbox{sgn}(y_i - y_j),
\end{equation}
where $\mbox{sgn}$ is the sign function, and $n$ is the number 
of data points in each data set. Kendall's $\tau$ takes values in the range $\tau \in [-1,+1]$.
A value of $\tau = \pm 1$ implies perfect agreement/disagreement between 
the two rankings. A value of $\tau = 0$ implies no relationship. 
We choose $\tau$ because it is non-parametric --- it measures the degree to which
a relationship between two parameters that describe a
data set results in the same ordering, without making assumptions about the functional form
of the relationship. This may be contrasted with the product-moment (Pearson) correlation 
coefficient, $r$, which measures the departure from a linear
relationship. 

A basic test of the significance of $\tau$ is to compute the $P$ 
value under the null hypothesis that there is no relationship
between the two rankings. Under the null hypothesis, 
the probability distribution for $\tau$ is
given exactly by  \citep{kendall1962rank}
\begin{equation}
  P_{H_0}(\tau, n) = \frac{F_n[S(\tau)]}{n!},
  \label{equ_tau_distro1}
\end{equation}
where
\begin{equation}
  S(\tau) = \frac{n(n-1)}{2} \tau
\end{equation}
and $F_n$ is defined by the recursion
\begin{equation}
  F_{n+1}(S) = \sum_{k=0}^{n} F_n(S+n-2k),
\end{equation}
with $F_n = 0$  whenever $S \notin [-n(n-1)/2,+n(n-1)/2]$. 
The recursion for $F$ is initiated with 
\begin{equation}
  F(S,2) = 1.
\end{equation}
It follows from the definition of $\tau$ that $S$ is always an integer, and $P_{H_0}$
is a discrete probability distribution. 
The two-sided $P$ value is the probability of 
obtaining a value of $|\tau|$ greater than or equal to the observed
value ($\tau_{\rm obs}$) under the null hypothesis. This value
is computed from Equation (\ref{equ_tau_distro1}) as 
\begin{equation}
  P = \sum_{S \ge S_{\rm obs}} P_{H_0}(S,n) + \sum_{S \le S_{\rm obs}} P_{H_0}(S,n),
\end{equation}
where $S_{\rm obs} =  S(\tau_{\rm obs})	$.
We have written a Python function for computing this distribution.
We have made our module available online \citep{DVN/NUWMFN_2020}.

The first four columns of Table \ref{table_tau} show Kendall's 
$\tau$ computed between the bin factor and each of the four 
metrics: $\aff$, $\maff$, $L_Q$, and $\emix$. We note that some entries
in Table \ref{table_tau} are exactly $\pm 1.00$. This
occurs because the data are perfectly monotonic and is not a result
of rounding to a finite precision in the table. If we were dealing
with the product-moment correlation coefficient, $r$, then finding
exactly $r = \pm 1$ for real-world (noisy) data would be cause
for some suspicion. However, for a given sample size, $n$, $\tau$ takes one of 
$1+n(n-1)/2$ rational values and will be exactly $\pm 1$ when the data
are perfectly monotonic.

Table \ref{table_pvals} shows the log of the two-sided $P$ value for the values 
of $\tau$ in Table \ref{table_tau}. A small value for $P$ indicates that the probability of 
obtaining the observed value of $\tau$ by chance is small. 
A large value indicates the opposite. In the following
we consider a $P$ value of $0.05$ as the threshold for significance. This
is a historically popular, but ultimately arbitrary, choice. 

Computing rank correlations in the presence of ties is more complicated than in the absence
of ties. In this context, a tie occurs when a metric has the same value for different bin
factors for a given method. Where necessary, we quote values in Table \ref{table_metrics} to sufficient precision to 
prevent the appearance of apparent ties due to rounding. Since we do not compute $\emix$, we are limited
to the precision quoted in Table 4 of \citet{2015ApJ...811..107D} for this metric. At this precision, a tie occurs in $\emix$ for bins
5 and 6 for FEMQ-N. We break the tie by adding a factor of either $-10^{-9}$ or $+10^{-9}$ to bin 4 of FEMQ-N. 
This gives values of $\tau$ equal to $.33$ and $.39$ respectively with an average of $0.36$. The corresponding $\log_{10}(P)$ 
values are $-0.59$ to $-0.74$ with an average of $-0.67$. The tie breaking does not significantly affect the results. 
In Tables \ref{table_tau} and \ref{table_pvals} we show the ``best case'' value, i.e. the largest $\tau$ value with the smallest $P$ value.

We consider the difference between $\aff$ and $\maff$. This
is an important comparison because it is a measure of the role
that the scaling problem plays in the trends noted by
\citet{2015ApJ...811..107D}.  For the metric $\aff$ we find $\tau_{\aff}$ 
close to unity in each case. However, for $\maff$ the trends are more complicated. 
Generally, $\tau_{\maff}$ is smaller in all cases.
MAGF achieves the smallest magnitude value of $\tau_{\maff}$, which is not significant 
based on the corresponding $P$ value, suggesting that there is no trend with spatial resolution for MAGF. 
In the case of CFIT, we find a significant negative value of $\tau_{\maff}$
for both the $P$ and $N$ solutions, indicating worse performance with increasing spatial resolution. Thus for some methods, the 
improvement with resolution reported by \citet{2015ApJ...811..107D} was likely due to the intrinsic scaling of $\aff$, 
but for more than half the methods, there is still a significant trend of improvement with resolution. Every
method showed some significant improvement with resolution (i.e. $\tau > 0$ with $P < 0.05$) for at least one metric.

For the $L_Q$ metric we find a lot of variation between methods.
CFIT-N, CFIT-P and OPTI became worse with increasing resolution. Some have 
$P$ values above a $0.05$ threshold. The amount of variation between methods as 
measured by the $L_Q$ metric is similar to that measured by $\maff$: CFIT again 
shows a significant worsening with resolution. In this case, OPTI also worsens with 
resolution, although the result is not statistically significant, while all the 
other methods show significant improvement with resolution.
It is important to note that $L_Q$ measures $\nabla\cdot\mathbf B$ in absolute terms and therefore
will tend to scale with the magnitude of $\mathbf B$. If one replaces
$\mathbf B$ by $\lambda \mathbf B$, where $\lambda$ is a constant,
then $L_Q$ becomes $\lambda L_Q$. The other metrics are normalized in
some sense and do not have this particular scaling. It is difficult
then to compare $L_Q$ between resolutions because the scale of $\mathbf B$ 
varies with bin factors. Indeed, \citet{2015ApJ...811..107D} discuss the effect of the binning on the 
inferred field strengths, the vertical electric current density $J_z$, and the total magnetic flux.

The metric $\emix$ has the fewest significant results. The methods MAGF and FEMQ do not 
achieve $P$ values below a $0.05$ threshold, suggesting no significant improvement with resolution.
For those methods where $\tau$ is significant, the trends are opposite those of $L_Q$, except for  
XTRAPOL. So, for example, CFIT shows improvement with spatial resolution by this metric. 

It is important to recall that $\tau$ measures monotonicity of data. 
It does not measure the strength of a particular relationship in 
absolute terms. Weakly varying data may be monotonic and have $\tau = 1$, but may also be
practically constant when measured in absolute terms. For example,
if two data sets, $x$ and $y$, are related by the linear relation
$y = \epsilon x + b$, where $\epsilon x \ll b$ and $b$ is a constant for the range of $x$
considered, then one finds $\tau = 1$ for these data. However, 
in absolute terms $y \approx b$. In this way, a value of $\tau$
close to zero is more informative as it indicates that no monotonic trend
exists either in terms of rank or in absolute terms of the data.
A value of $\tau \pm 1$ indicates a strong correlation in rank, but
the data may vary little when considered in absolute terms.

\subsection{Measure of concordance between different metrics}
\label{sec_concord}

To measure the agreement/disagreement between different metrics,
we compute Kendall's coefficient of concordance, $W$, for 
three of the metrics \citep{kendall1962rank,daniel1978applied}.

A set of $n$ ``objects'' can be ranked in $m$ different ways according
to different metrics. If we define $v_{ij}$  as the rank of 
object $i$ according to ranking $j$, then the coefficient 
of concordance is defined as \citep{kendall1962rank}
\begin{equation}
  W = \frac{12}{m^2 (n^3 - n)} \sum_{i=1}^n (u_i - \overline{u})^2, 
  \label{equ_W_def}
\end{equation}
where
\begin{equation}
  u_i = \sum_{j=1}^m v_{ij}
  \label{equ_W_def2}
\end{equation}
is the sum of the ranks over the different rankings, and
\begin{equation}
  \overline{u} = \frac{1}{2} m (n-1).
\end{equation}
Kendall's $W$ measures the extent to which the $m$ rankings agree. 
It takes a value in the range $[0,1]$. A value of $W=1$ indicates perfect
agreement between the $m$ rankings. A value of $W=0$ indicates
no agreement. The $P$ value for $W$
under the null hypothesis can be computed from the asymptotic formula 
\citep{kendall1962rank}.
\begin{equation}
  P = P_{\chi^2}[m(n-1)W],
\end{equation}
where $P_{\chi^2}$ is the $\chi^2$ distribution with $n-1$ degrees of freedom.
When computing, $W$ we use the ``correction for continuity'' described by \citet{kendall1962rank}, 
which is appropriate for small sample sizes. The correction is performed by subtracting
one from the numerator and adding two to the denominator of the
ratio in Equation (\ref{equ_W_def}). It should be noted that the 
form of $W$ and the corresponding $P$ value are 
only appropriate when there are no ties in the data. More complex expressions
are required when ties are present \citep{kendall1962rank}.
We have developed a Python module for 
evaluating both $W$ and the asymptotic $P$ value. This module
utilizes basic numerical functions from the SciPy library \citep{2020SciPy-NMeth}.
We have made our module available online \citep{DVN/NUWMFN_2020}.

In the present context, the ``objects'' are the NLFFF solutions at 
different resolutions for a give method, and the metrics are those that 
we have defined in Sections \ref{sec_metric} and \ref{sec_other_metrics}.
We consider three metrics, so $m=3$ in our case, and $n$ is the number 
of different bin factors: $n=8$ for MAGF and $n=9$ for all
the other methods. 

The final column of Table \ref{table_tau} shows $W$ for $\maff$, 
$L_Q$, and $\emix$. It measures the agreement between these
three metrics. We exclude $\aff$ from the calculation, because of
the scaling problem. The final column of Table \ref{table_pvals}
shows the $\log_{10}$ of the $P$ values for $W$ for each of the 
codes/methods. 

No method achieves a perfect score of $W=1$, although
XTRAPOL and FEMQ come the closest. The other methods generally achieve
values of $W < 0.4$ and are not significant, according to their $P$ values.
Only XTRAPOL and FEMQ have $P$ values below a $0.05$ level of significance, 
suggesting that for the other metrics there is no real association
between the rankings given by $\maff$, $L_Q$, and $\emix$. 

\section{Discussion and Conclusions}
\label{sec_con}

The metric $\aff$ as originally defined by
\citet{2000ApJ...540.1150W} is problematic as a metric 
for measuring divergence because
it exhibits a scaling problem: $\aff \sim (\Delta V)^{1/3}$
regardless of $\nabla\cdot\mathbf B$,  where here $\Delta V$ is the volume of a mesh cell
on which $\mathbf B$ is defined.
This means that comparing $\aff$ computed on different
meshes is ill-advised, because $\aff$ naturally becomes smaller when using
a finer mesh, even without any actual change in $\nabla\cdot\mathbf B$. 
To address this deficiency, we define a new metric, $\maff$, which is a simple modification of $\aff$. 
As shown in Section \ref{sec_metric}, the new metric has the improved scaling $\maff \sim c + \mathcal{O}(\Delta V^{1/3})$,
where $c$ is independent of $\Delta V$. Hence, $\maff$ is
not asymptotic to zero for small $\Delta V$ as $\aff$ is.

We also revisit the issue considered by \citet{2015ApJ...811..107D}
of whether spatial resolution affects $\nabla\cdot\mathbf B$ for 
NLFFF extrapolations. We consider the two divergence metrics 
computed by \citet{2015ApJ...811..107D}, i.e. $\aff$ and $\emix$.
We also consider $\maff$ and $\Lq$. Our aims are threefold.
First we aim to assess the effect of the scaling problem for $\aff$
on the results \citet{2015ApJ...811..107D}. Second, we aim to 
perform a quantitative analysis of the trends in spatial resolution
for the four metrics. Third, we aim to measure the concordance between different metrics, 
i.e. if we rank solutions using different metrics, to what extent
do these rankings agree/differ.

In Section \ref{sec_rank} we compute Kendall's rank-correlation coefficient, $\tau$,
for the different metrics/methods. The trends are more complicated than those
reported in \citet{2015ApJ...811..107D}, suggesting that the scaling problem for $\aff$ 
is partially responsible for those results.
From our results, it appears that some NLFFF methods perform
worse than others in terms of satisfying the $\nabla\cdot\mathbf B =0$
condition. XTRAPOL and FEMQ have the smallest magnitude of each $\nabla\cdot\mathbf B$ 
metric at almost every spatial resolution, but FEMQ does not have a consistent 
trend of decreasing $\emix$ with increasing spatial resolution ($\tau=0.39$ with $\log_{10}(P)=-0.74$). On the other hand,  
the magnetofrictional method (MAGF) typically has the largest magnitude for each metric. For this method,
only the metric $L_Q$ appears to improve significantly with resolution. 
The results for CFIT and the optimization method (OPTI) are mixed. For some
of the metrics they have $\tau < 0$
indicating increasing $|\nabla\cdot\mathbf B|$ with resolution.

From our analysis, it would appear that some NLFFF solution methods are worse than others in 
terms of achieving $\nabla\cdot\mathbf B =0$.
However, some caution is required
when drawing conclusions of this nature. The \citet{2015ApJ...811..107D}
results depend not only on the NLFFF method used, but also on the 
various ways the boundary data were treated.
As described in \citet{2015ApJ...811..107D}, the binned boundary
data were smoothed, censored, and preprocessed in different ways
depending on the NLFFF method used. It is difficult, therefore, to 
completely separate the effects of the NLFFF method from the effects
of the processing. 

Although the processing methods are different, we expect a general
reduction in $|\nabla\cdot\mathbf B|$ as 
electric currents are removed from the boundary data due to smoothing/censoring.
As electric current is removed, the NLFFF solution approaches a
potential field. The construction of a potential field is a well-posed mathematical
problem that is straightforward to implement numerically, and we expect 
negligible $\nabla\cdot\mathbf B$ violations for this special case. 
We therefore expect a general reduction in $|\nabla\cdot\mathbf B|$ for 
NLFFF solutions as the limit of a potential field is approached.

As discussed in Section \ref{sec_concord}, we find values for the coefficient of concordance, $W$, 
that are statistically consistent with zero for all but two methods (FEMQ and XTRAPOL). This indicates
that the ranking of solutions from best to worst generally depends on the metric. In particular, 
in some instances, what is regarded as the most solenoidal solution depends on the choice of metric.

In the limit that $\nabla\cdot\mathbf B$ goes to zero, one expects
some association between the metrics $\maff$, $L_Q$, and $\emix$. 
However, for finite $\nabla\cdot\mathbf B$, these metrics may differ because
they depend on the distribution of $\nabla\cdot\mathbf B$
and $\mathbf B$ in different ways. The $\maff$ metric is normalized
by $|\mathbf B|$, whereas $L_Q$ is not. The metric $\emix$ depends 
not just on $\nabla\cdot\mathbf B$,  but also on the orientation of the 
non-solenoidal field relative to the solenoidal field \citep{2013A&A...553A..38V}.
Both \citet{2013A&A...553A..38V} and \citet{2015ApJ...811..107D} found
that the $\aff$ metric does not predict $\emix$. Given these results,
we conclude that there is value in computing different metrics
for $\nabla\cdot\mathbf B$.

For the metrics that we consider, a smaller value is better in the sense
that it indicates a more divergence-free magnetic field. How small, then,
do these metrics need to be before an NLFFF solution should be accepted? 
In some contexts this question has a definitive answer. For example, 
to use an NLFFF extrapolation to estimate free energy, a common application,
it is necessary that $\emix$ and the other non-solenoidal energy components be 
smaller than the measured free energy, otherwise the free energy is unphysical. 
In other contexts, the answer is unclear. How large do the metrics need to
be before either the helicity or topology of an NLFFF extrapolation becomes unreliable?
A priori, the answer to this question is unclear, and more research is required to
properly address it. As a first step, we recommend the reporting of these metrics so that it is at least possible to make comparisons 
between different studies. 

We acknowledge that in both our approach and that of \citet{2015ApJ...811..107D} 
$\nabla\cdot\mathbf B$ is computed using a method that is inconsistent
with the way derivatives are approximated by the NLFFF codes. 
We compute $\nabla\cdot\mathbf B$ using a second-order finite-difference
approximation to the derivatives, whereas the NLFFF codes
use a variety of alternatives. For example, FEMQ
is a finite element code, and CFIT is based on a Fourier spectral
method. In using a method of numerical differentiation that differs
from the codes some additional truncation error is introduced, and thus
our analysis reflects trends in not only the codes/methods, but 
also the truncation error introduced in computing $\nabla\cdot\mathbf B$ itself.

The reliability of our statistical approach may also be questioned
given the small number of data points under consideration. We compute
$\tau$ and the $P$ values from eignt to nine data points in each case. 
How reliable are these numbers? We can be
confident that the $P$ values for $\tau$ are meaningful because the probability 
distribution for $\tau$ under the null hypothesis can be computed 
exactly for any sample size \citep{kendall1962rank} --- we do not
rely on a large sample size to justify assumptions of asymptotic
normality in deriving $P$ for $\tau$, for example. As noted previously, $P$ 
only measures the significance of $\tau$ from zero. We have not 
computed confidence intervals for $\tau$, which is nontrivial
given the small data set. For the calculation of the $P$ values for $W$,
we rely on an asymptotic distribution. However, \citet{kendall1962rank}, recommends
this approach for a sample size of $n > 7$. Hence, the $P$ values computed in this
way are unlikely to be significantly different from those computed from an exact
distribution for the null hypothesis of $W$.

In summary, we have shown that the average fractional flux $\aff$ is generally
a poor measure of the divergence due to an intrinsic scaling problem
and should be replaced by the modified fractional flux $\maff$. 
In re-analyzing the results of \citet{2015ApJ...811..107D} we find
that the scaling problem masks a more complicated trend. More generally,
we find that measuring divergence depends somewhat on how it is being
measured: different metrics may give different results. Therefore,
it is recommended to calculate more than one metric. As NLFFF extrapolations
are used often, it is increasingly important to quantify 
$\nabla\cdot\mathbf B = 0$ violations in order to 
meaningfully interpret the results of these calculations.

\acknowledgments

This material is based upon work supported by the National Science 
Foundation under Grant Nos.\ 1841962 and 1630454, and by 
NASA award No.\ 80NSSC18K0071. Any opinions, findings, and 
conclusions or recommendations expressed in this material are those of 
the authors and do not necessarily reflect the views of either the National 
Science Foundation or the National Aeronautics and Space Administration.
The research presented in this article is based on data resulting from the
meetings of International Team 238, ``Nonlinear Force-Free Modeling of the Solar Corona: Toward 
a New Generation of Methods,'' held in 2013 and 2014 at the 
International Space Science Institute (ISSI) in Bern, Switzerland.


\pagebreak
%
%
%
%
%

\pagebreak

\newpage
\startlongtable
\begin{deluxetable}{llllll}
\tablecaption{Divergence metrics for extrapolations of \citet{2015ApJ...811..107D}.
The labels indicate the method used to compute the NLFFF extrapolation 
and are explained in the text. For each method, several extrapolations
were performed at different spatial resolutions. The bin factor indicates
the factor by which the {\it Hinode}/SOT data were binned before deriving
boundary conditions for the NLFFF modeling. We compute the values
in  columns 3-5 ourselves. The values in column 6 are reproduced
from Table 4 of \citet{2015ApJ...811..107D}. The results in the last four 
columns are plotted in Figure \ref{fig_metrics} versus bin factor.\label{table_metrics}}
\tablehead{
  \colhead{Method} & 
  \colhead{Bin}    & 
  \colhead{$\aff$[$\times 10^{-4}$]} & 
  \colhead{$\maff$[$\times 10^{-9}\,{\rm m^{-1}}$]} &
  \colhead{$L_Q$[$\times 10^{-8}\,{\rm Mx/cm^{3}}$]} & 
  \colhead{$\emix$[ $\times 10^{-2}$]} 
}
\startdata
     CFIT (N/P) & 02 & 3.39/2.13 & 9.60/6.03 & 2.15/1.71 & 0.05/0.05 \\ 
                & 03 & 4.78/3.99 & 9.02/7.54 & 2.11/1.90 & 0.09/0.09 \\ 
                & 04 & 6.11/4.96 & 8.65/7.02 & 2.17/2.01 & 2.55/1.84 \\ 
                & 06 & 6.13/4.69 & 5.79/4.43 & 1.81/1.65 & 0.24/0.25 \\ 
                & 08 & 7.27/6.35 & 5.15/4.50 & 1.68/1.59 & 0.36/0.39 \\ 
                & 10 & 8.64/6.53 & 4.89/3.70 & 1.70/1.48 & 0.51/0.53 \\ 
                & 12 & 9.81/7.56 & 4.63/3.57 & 1.62/1.44 & 0.64/0.61 \\ 
                & 14 & 11.4/9.86 & 4.61/3.99 & 1.60/1.54 & 0.90/0.88 \\ 
                & 16 & 11.9/11.1 & 4.21/3.93 & 1.54/1.50 & 1.07/1.02 \\ 
     \hline
FEMQ (N/P) & 02  & 0.387/0.424 & 1.10/1.20   & 0.516/0.541 & 0.024/0.023 \\ 
           & 03  & 0.663/0.695 & 1.25/1.31   & 0.701/0.714 & 0.066/0.064 \\ 
           & 04  & 0.981/0.988 & 1.39/1.40   & 0.750/0.748 & 0.001/0.003 \\ 
           & 06  & 1.75/1.77   & 1.65/1.67   & 0.856/0.861 & 0.001/0.001 \\ 
           & 08  & 2.78/2.78   & 1.97/1.97   & 1.05/1.05   & 0.275/0.272 \\ 
           & 10  & 3.52/3.50   & 1.99/1.98   & 1.00/1.00   & 0.199/0.183 \\ 
           & 12  & 4.61/4.51   & 2.18/2.13   & 1.03/1.02   & 0.197/0.205 \\ 
           & 14  & 5.30/5.23   & 2.14/2.12   & 1.06/1.06   & 0.461/0.454 \\ 
           & 16  & 6.86/6.40   & 2.43/2.27   & 1.07/1.05   & 0.120/0.117 \\ 
     \hline
    MAGF & 02 & 13.0 & 36.8 & 3.16 & 0.26 \\ 
         & 04 & 13.3 & 18.8 & 4.47 & 2.58 \\ 
         & 06 & 17.5 & 16.5 & 4.02 & 3.02 \\ 
         & 08 & 32.9 & 23.3 & 4.60 & 2.12 \\ 
         & 10 & 46.4 & 26.3 & 5.25 & 2.33 \\ 
         & 12 & 51.6 & 24.4 & 5.30 & 0.68 \\   
         & 14 & 70.9 & 28.7 & 4.66 & 0.65 \\ 
         & 16 & 82.4 & 29.2 & 5.09 & 0.22 \\ 
     \hline
    OPTI & 02 & 1.14 & 3.23 & 1.40  & 11.0 \\      
         & 03 & 1.75 & 3.30 & 1.51  & 7.70 \\      
         & 04 & 2.18 & 3.09 & 1.14  & 9.75 \\ 
         & 06 & 3.70 & 3.49 & 1.20  & 7.13 \\ 
         & 08 & 6.14 & 4.35 & 1.09  & 8.49 \\ 
         & 10 & 10.2 & 5.76 & 1.324 & 6.63 \\ 
         & 12 & 15.7 & 7.41 & 1.316 & 9.21 \\ 
         & 14 & 17.7 & 7.16 & 1.24  & 5.88 \\ 
         & 16 & 19.5 & 6.90 & 1.04  & 5.75 \\ 
     \hline
    XTRA (N/P) & 02 & 0.403/0.46 & 1.14/1.30 & 0.592/0.669  & 0.050/0.048 \\   
               & 03 & 0.77/0.828 & 1.45/1.56 & 0.763/0.806  & 0.106/0.101 \\       
               & 04 & 1.20/1.24  & 1.70/1.75 & 0.904/0.925  & 0.149/0.146 \\ 
               & 06 & 2.11/2.20  & 1.99/2.08 & 1.03/1.06    & 0.275/0.281 \\ 
               & 08 & 3.05/3.07  & 2.16/2.18 & 1.11/1.12    & 0.389/0.393 \\ 
               & 10 & 4.69/5.06  & 2.66/2.86 & 1.24/1.26    & 0.575/0.560 \\ 
               & 12 & 5.67/6.00  & 2.68/2.83 & 1.21372/1.24 & 0.615/0.603 \\ 
               & 14 & 6.49/6.71  & 2.63/2.72 & 1.21371/1.22 & 0.936/0.890 \\ 
               & 16 & 7.85/7.85  & 2.78/2.78 & 1.27/1.29    & 1.05/1.01 \\ 
\enddata
\end{deluxetable}

%
%
\newpage
\begin{table}
\begin{tabular}{llllll}
     Method & $\tau_{\aff}$ & $\tau_{\maff}$ & $\tau_{L_Q}$ & $\tau_{\emix}$ & W \\
     \hline
     CFIT-N & 1.00 & -1.00 & -0.83 & 0.67 & 0.21 \\ 
     CFIT-P & 0.94 & -0.61 & -0.61 & 0.67 & 0.30 \\ 
     FEMQ-N & 1.00 & 0.94 & 0.89 & 0.39 & 0.83 \\    
     FEMQ-P & 1.00 & 0.94 & 0.83 & 0.39 & 0.81 \\ 
     MAGF & 1.00 & 0.36 & 0.64 & -0.43 & 0.34 \\ 
     OPTI & 1.00 & 0.72 & -0.39 & -0.61 & 0.28 \\ 
     XTRA-N & 1.00 & 0.89 & 0.83 & 1.00 & 0.97 \\ 
     XTRA-P & 1.00 & 0.72 & 0.83 & 1.00 & 0.92 \\      
\end{tabular}
 \caption{The first four columns give the Kendall $\tau$ 
   rank-correlation between bin size and the four metrics
   $\aff$, $\maff$, $L_Q$, and $\emix$ computed from the 
   \citet{2009ApJ...696.1780D} solution data values. A value of $\tau = 1$
   indicates that the particular metric is monotonically increasing
   with bin factor (and therefore monotonically decreasing with 
   increasing resolution). A value of $\tau = -1$ indicates the opposite.   
   The final column contains the coefficient of concordance, $W$, 
   described in Section \ref{sec_concord} for three of the four metrics.}
 \label{table_tau}
\end{table}

%
%
\newpage
\begin{table}
  \begin{tabular}{llllll}
      Method & $P(\tau_{\aff})$ & $P(\tau_{\maff})$ & $P(\tau_{L_Q})$ & $P(\tau_{\emix})$ & $P(W)$ \\ 
       \hline
      CFIT-N & -5.26 & -5.26 & -3.07 & -1.90 & -0.13 \\ 
       CFIT-P & -4.30 & -1.61 & -1.61 & -1.90 & -0.28 \\ 
       FEMQ-N & -5.26 & -4.30 & -3.62 & -0.74 & -1.98 \\ 
       FEMQ-P & -5.26 & -4.30 & -3.07 & -0.74 & -1.88 \\ 
       MAGF & -4.30 & -0.56 & -1.51 & -0.75 & -0.37 \\ 
       OPTI & -5.26 & -2.23 & -0.74 & -1.61 & -0.25 \\ 
       XTRA-N & -5.26 & -3.62 & -3.07 & -5.26 & -2.51 \\ 
       XTRA-P & -5.26 & -2.23 & -3.07 & -5.26 & -2.34 \\ 
   \end{tabular}
   \caption{Table of the $\log_{10}(P)$ values for the $\tau$ and $W$ results
   in Table \ref{table_tau}. The $P$-value is the probability
   of obtaining the results or a more extreme value in Table \ref{table_tau} under the null
   hypotheses. The null hypotheses are explained in
   Sections \ref{sec_rank} and \ref{sec_concord}.   
   For reference, the $0.05$ significance level in the
   log scale is $\log_{10}(0.05) \approx -1.30$.}
   \label{table_pvals}   
\end{table}

\pagebreak


\begin{thebibliography}{}
\expandafter\ifx\csname natexlab\endcsname\relax\def\natexlab#1{#1}\fi
\providecommand{\url}[1]{\href{#1}{#1}}
\providecommand{\dodoi}[1]{doi:~\href{http://doi.org/#1}{\nolinkurl{#1}}}
\providecommand{\doeprint}[1]{\href{http://ascl.net/#1}{\nolinkurl{http://ascl.net/#1}}}
\providecommand{\doarXiv}[1]{\href{https://arxiv.org/abs/#1}{\nolinkurl{https://arxiv.org/abs/#1}}}

\bibitem[{{Aly}(1989)}]{1989SoPh..120...19A}
{Aly}, J.~J. 1989, \solphys, 120, 19, \dodoi{10.1007/BF00148533}

\bibitem[{{Amari} \& {Aly}(2010)}]{2010A&A...522A..52A}
{Amari}, T., \& {Aly}, J.-J. 2010, \aap, 522, A52,
  \dodoi{10.1051/0004-6361/200913058}

\bibitem[{{Amari} {et~al.}(2006){Amari}, {Boulmezaoud}, \&
  {Aly}}]{2006A&A...446..691A}
{Amari}, T., {Boulmezaoud}, T.~Z., \& {Aly}, J.~J. 2006, \aap, 446, 691,
  \dodoi{10.1051/0004-6361:20054076}

\bibitem[{Daniel(1978)}]{daniel1978applied}
Daniel, W. 1978, Applied nonparametric statistics, 1st edn. (Boston,MA,USA:
  Houghton Mifflin Company), 298

\bibitem[{{DeRosa} {et~al.}(2009){DeRosa}, {Schrijver}, {Barnes}, {Leka},
  {Lites}, {Aschwanden}, {Amari}, {Canou}, {McTiernan}, {R{\'e}gnier},
  {Thalmann}, {Valori}, {Wheatland}, {Wiegelmann}, {Cheung}, {Conlon},
  {Fuhrmann}, {Inhester}, \& {Tadesse}}]{2009ApJ...696.1780D}
{DeRosa}, M.~L., {Schrijver}, C.~J., {Barnes}, G., {et~al.} 2009, \apj, 696,
  1780, \dodoi{10.1088/0004-637X/696/2/1780}

\bibitem[{{DeRosa} {et~al.}(2015){DeRosa}, {Wheatland}, {Leka}, {Barnes},
  {Amari}, {Canou}, {Gilchrist}, {Thalmann}, {Valori}, {Wiegelmann},
  {Schrijver}, {Malanushenko}, {Sun}, \& {R{\'e}gnier}}]{2015ApJ...811..107D}
{DeRosa}, M.~L., {Wheatland}, M.~S., {Leka}, K.~D., {et~al.} 2015, \apj, 811,
  107, \dodoi{10.1088/0004-637X/811/2/107}

\bibitem[{{Fan} {et~al.}(2012){Fan}, {Wang}, {He}, \&
  {Zhu}}]{2012RAA....12..563F}
{Fan}, Y.-L., {Wang}, H.-N., {He}, H., \& {Zhu}, X.-S. 2012, Research in
  Astronomy and Astrophysics, 12, 563, \dodoi{10.1088/1674-4527/12/5/008}

\bibitem[{{Gary}(2001)}]{2001SoPh..203...71G}
{Gary}, G.~A. 2001, \solphys, 203, 71, \dodoi{10.1023/A:1012722021820}

\bibitem[{{Gilchrist}(2020)}]{DVN/NUWMFN_2020}
{Gilchrist}, S.~A. 2020, {Python Rank Stat. codes}, Version 1.0,  Harvard
  Dataverse, \dodoi{10.7910/DVN/NUWMFN}

\bibitem[{Grad \& Rubin(1958)}]{24756}
Grad, H., \& Rubin, H. 1958, in Peaceful Uses of Atomic Energy: Theoretical and
  Experimental Aspects of Controlled Nuclear Fusion, ed. J.~H. Martens,
  L.~Ourom, W.~M. Barss, L.~G. Bassett, K.~R.~E. Smith, M.~Gerrard,
  F.~Hudswell, B.~Guttman, J.~H. Pomeroy, W.~B. Woollen, K.~S. Singwi, T.~E.~F.
  Carr, A.~C. Kolb, A.~H.~S. Matterson, S.~P. Welgos, I.~D. Rojanski, \&
  D.~Finkelstein, Vol.~31 (Geneva: United Nations), 190--197

\bibitem[{Kendall(1962)}]{kendall1962rank}
Kendall, M. 1962, Rank correlation methods, 3rd edn. (Hafner Pub. Co.)

\bibitem[{{Lites} {et~al.}(2013){Lites}, {Akin}, {Card}, {Cruz}, {Duncan},
  {Edwards}, {Elmore}, {Hoffmann}, {Katsukawa}, {Katz}, {Kubo}, {Ichimoto},
  {Shimizu}, {Shine}, {Streander}, {Suematsu}, {Tarbell}, {Title}, \&
  {Tsuneta}}]{2013SoPh..283..579L}
{Lites}, B.~W., {Akin}, D.~L., {Card}, G., {et~al.} 2013, \solphys, 283, 579,
  \dodoi{10.1007/s11207-012-0206-3}

\bibitem[{{Mastrano} {et~al.}(2018){Mastrano}, {Wheatland}, \&
  {Gilchrist}}]{2018SoPh..293..130M}
{Mastrano}, A., {Wheatland}, M.~S., \& {Gilchrist}, S.~A. 2018, \solphys, 293,
  130, \dodoi{10.1007/s11207-018-1351-0}

\bibitem[{{Metcalf} {et~al.}(1995){Metcalf}, {Jiao}, {McClymont}, {Canfield},
  \& {Uitenbroek}}]{1995ApJ...439..474M}
{Metcalf}, T.~R., {Jiao}, L., {McClymont}, A.~N., {Canfield}, R.~C., \&
  {Uitenbroek}, H. 1995, \apj, 439, 474, \dodoi{10.1086/175188}

\bibitem[{{Moraitis} {et~al.}(2014){Moraitis}, {Tziotziou}, {Georgoulis}, \&
  {Archontis}}]{2014SoPh..289.4453M}
{Moraitis}, K., {Tziotziou}, K., {Georgoulis}, M.~K., \& {Archontis}, V. 2014,
  \solphys, 289, 4453, \dodoi{10.1007/s11207-014-0590-y}

\bibitem[{Press {et~al.}(2007)Press, Teukolsky, Vetterling, \&
  Flannery}]{Press:2007:NRE:1403886}
Press, W.~H., Teukolsky, S.~A., Vetterling, W.~T., \& Flannery, B.~P. 2007,
  Numerical Recipes: 3rd Edition: The Art of Scientific Computing, 3rd edn. (New
  York, NY, USA: Cambridge University Press)

\bibitem[{{R{\'e}gnier}(2013)}]{2013SoPh..288..481R}
{R{\'e}gnier}, S. 2013, \solphys, 288, 481, \dodoi{10.1007/s11207-013-0367-8}

\bibitem[{{Sakurai}(1989)}]{1989SSRv...51...11S}
{Sakurai}, T. 1989, \ssr, 51, 11, \dodoi{10.1007/BF00226267}

\bibitem[{{Schrijver} {et~al.}(2006){Schrijver}, {De Rosa}, {Metcalf}, {Liu},
  {McTiernan}, {R{\'e}gnier}, {Valori}, {Wheatland}, \&
  {Wiegelmann}}]{2006SoPh..235..161S}
{Schrijver}, C.~J., {De Rosa}, M.~L., {Metcalf}, T.~R., {et~al.} 2006,
  \solphys, 235, 161, \dodoi{10.1007/s11207-006-0068-7}

\bibitem[{Sturrock \& Andrew(1994)}]{sturrock1994plasma}
Sturrock, P., \& Andrew, S. 1994, Plasma Physics: An Introduction to the Theory
  of Astrophysical, Geophysical and Laboratory Plasmas, Stanford-Cambridge
  program (Cambridge University Press)

\bibitem[{{Su} {et~al.}(2014){Su}, {Jing}, {Wang}, {Wiegelmann}, \&
  {Wang}}]{2014ApJ...788..150S}
{Su}, J.~T., {Jing}, J., {Wang}, S., {Wiegelmann}, T., \& {Wang}, H.~M. 2014,
  \apj, 788, 150, \dodoi{10.1088/0004-637X/788/2/150}

\bibitem[{{Thalmann} {et~al.}(2012){Thalmann}, {Pietarila}, {Sun}, \&
  {Wiegelmann}}]{2012AJ....144...33T}
{Thalmann}, J.~K., {Pietarila}, A., {Sun}, X., \& {Wiegelmann}, T. 2012, \aj,
  144, 33, \dodoi{10.1088/0004-6256/144/2/33}

\bibitem[{{Thalmann} \& {Wiegelmann}(2008)}]{2008A&A...484..495T}
{Thalmann}, J.~K., \& {Wiegelmann}, T. 2008, \aap, 484, 495,
  \dodoi{10.1051/0004-6361:200809508}

\bibitem[{{Tsuneta} {et~al.}(2008){Tsuneta}, {Ichimoto}, {Katsukawa}, {Nagata},
  {Otsubo}, {Shimizu}, {Suematsu}, {Nakagiri}, {Noguchi}, {Tarbell}, {Title},
  {Shine}, {Rosenberg}, {Hoffmann}, {Jurcevich}, {Kushner}, {Levay}, {Lites},
  {Elmore}, {Matsushita}, {Kawaguchi}, {Saito}, {Mikami}, {Hill}, \&
  {Owens}}]{2008SoPh..249..167T}
{Tsuneta}, S., {Ichimoto}, K., {Katsukawa}, Y., {et~al.} 2008, \solphys, 249,
  167, \dodoi{10.1007/s11207-008-9174-z}

\bibitem[{{Valori} {et~al.}(2013){Valori}, {D{\'e}moulin}, {Pariat}, \&
  {Masson}}]{2013A&A...553A..38V}
{Valori}, G., {D{\'e}moulin}, P., {Pariat}, E., \& {Masson}, S. 2013, \aap,
  553, A38, \dodoi{10.1051/0004-6361/201220982}

\bibitem[{{Valori} {et~al.}(2007){Valori}, {Kliem}, \&
  {Fuhrmann}}]{2007SoPh..245..263V}
{Valori}, G., {Kliem}, B., \& {Fuhrmann}, M. 2007, \solphys, 245, 263,
  \dodoi{10.1007/s11207-007-9046-y}

\bibitem[{{Valori} {et~al.}(2010){Valori}, {Kliem}, {T{\"o}r{\"o}k}, \&
  {Titov}}]{2010A&A...519A..44V}
{Valori}, G., {Kliem}, B., {T{\"o}r{\"o}k}, T., \& {Titov}, V.~S. 2010, \aap,
  519, A44, \dodoi{10.1051/0004-6361/201014416}

\bibitem[{{Virtanen} {et~al.}(2020){Virtanen}, {Gommers}, {Oliphant},
  {Haberland}, {Reddy}, {Cournapeau}, {Burovski}, {Peterson}, {Weckesser},
  {Bright}, {van der Walt}, {Brett}, {Wilson}, {Jarrod Millman}, {Mayorov},
  {Nelson}, {Jones}, {Kern}, {Larson}, {Carey}, {Polat}, {Feng}, {Moore}, {Vand
  erPlas}, {Laxalde}, {Perktold}, {Cimrman}, {Henriksen}, {Quintero}, {Harris},
  {Archibald}, {Ribeiro}, {Pedregosa}, {van Mulbregt}, \&
  {Contributors}}]{2020SciPy-NMeth}
{Virtanen}, P., {Gommers}, R., {Oliphant}, T.~E., {et~al.} 2020, Nature
  Methods, 17, 261, \dodoi{https://doi.org/10.1038/s41592-019-0686-2}

\bibitem[{{Wheatland}(2007)}]{2007SoPh..245..251W}
{Wheatland}, M.~S. 2007, \solphys, 245, 251, \dodoi{10.1007/s11207-007-9054-y}

\bibitem[{{Wheatland} {et~al.}(2000){Wheatland}, {Sturrock}, \&
  {Roumeliotis}}]{2000ApJ...540.1150W}
{Wheatland}, M.~S., {Sturrock}, P.~A., \& {Roumeliotis}, G. 2000, \apj, 540,
  1150, \dodoi{10.1086/309355}

\bibitem[{{Wiegelmann} \& {Inhester}(2010)}]{2010A&A...516A.107W}
{Wiegelmann}, T., \& {Inhester}, B. 2010, \aap, 516, A107,
  \dodoi{10.1051/0004-6361/201014391}

\bibitem[{{Wiegelmann} \& {Sakurai}(2012)}]{2012LRSP....9....5W}
{Wiegelmann}, T., \& {Sakurai}, T. 2012, Living Reviews in Solar Physics, 9, 5,
  \dodoi{10.12942/lrsp-2012-5}

\bibitem[{{Wiegelmann} {et~al.}(2012){Wiegelmann}, {Thalmann}, {Inhester},
  {Tadesse}, {Sun}, \& {Hoeksema}}]{2012SoPh..281...37W}
{Wiegelmann}, T., {Thalmann}, J.~K., {Inhester}, B., {et~al.} 2012, \solphys,
  281, 37, \dodoi{10.1007/s11207-012-9966-z}

\end{thebibliography}
\end{document}